\documentclass{ws-p8-50x6-00}
\usepackage{latexsym,alltt}
\newcommand{\tomath}[1]{{\ifmmode{#1}\else${#1}$\fi}}
\def\frc#1#2{\frac{\displaystyle\strut #1}{\displaystyle\strut #2}}
\newcommand{\rl}[3]{\mbox{\sc #1}\ \frc{#2}{#3}}

\newcommand{\rlref}[1]{{\textsc{#1}}}
\newcommand{\mathspace}{\ifmmode\:\fi}  % genera uno spazio solo in modo matematico
\newcommand{\mc}[1]{{\normalfont\it #1}}
\newcommand{\cm}[1]{\texttt{#1}}   % i comandi sono in teletype
\newcommand{\ct}[1]{\textsl{#1}}   % le categorie
\newcommand{\Ra}{\Rightarrow}

\newcommand{\spc}{\hspace{1.2em}}

\newcommand{\defeq}{\stackrel{\rm def}{=}}

\newcommand{\vcr}{\mid\kern-.4ex\protect\nolinebreak\models}
\newcommand{\tcr}{\models}
\newcommand{\ftcr}{\models{_{\hspace{-1ex}f}}}
\newcommand{\vvdash}{\vdash\kern-1ex\protect\nolinebreak\vdash}
\newcommand{\vvvdash}{\vdash\kern-1ex\protect\nolinebreak\vvdash}

\newcommand{\Coq}{\textrm{\sf Coq}}
\newcommand{\uK}{\ensuremath{\mu K}}
\def\ELF+{\ensuremath{\textrm{ELF}^+}}

\newcommand{\fv}{{\rm F\kern-0.25emV}}
\newcommand{\ff}{{\rm F\kern-0.25emF}}
\newcommand{\fb}{{\rm F\kern-0.25emB}}
\newcommand{\fp}{{\rm F\kern-0.25emP}}
\newcommand{\fu}{{\rm F\kern-0.20emOp}}
\newcommand{\dv}{{\rm D\kern-0.27emV}}
\newcommand{\av}{{\rm A\kern-0.40emV}}

\newcommand{\In}{{\rm in}\kern-0.1em}
\newcommand{\Is}{{\rm is}\kern-0.1em}

\newcommand{\dsb}[1]{[\kern-0.14em[ #1 ]\kern-0.14em]}
\renewcommand{\phi}{\varphi}
\renewcommand{\epsilon}{\varepsilon}

\newcommand{\ps}{{\cal P}}      % simbolo per insieme potenza
\newcommand{\fs}{{\cal P}_{\kern-.3em <\omega}}      % simbolo per l'insieme di sottoinsiemi finiti
\newcommand{\res}{{\cal P}_{\rm \kern-.3em re}}      % simbolo per l'insieme di sottoinsiemi r.e.

\newcommand{\N}[1]{\protect\tomath{{\rm\bf N}#1}}

\newcommand{\PAQ}{\tomath{PA^{\rm \kern-0.1em qf}}}

\newcommand{\DL}{\ensuremath{\mathit{DL}}}
\newcommand{\PDL}{\ensuremath{\mathit{PDL}}}
\newcommand{\ADL}{\ensuremath{\DL^{\rm \kern-0.1em qf}}}
\newcommand{\M}{{\ensuremath{\cal M}}}

\newcommand{\ELSE}{\mbox{\rm [\kern-1pt]}\:}

\newcommand{\BOX}[2]{\left[ #1 \right] #2}

\newtheorem{theorem}{Theorem}

\newtheorem{definition}{Definition}
\newtheorem{proposition}[theorem]{Proposition}

\newenvironment{proof}
               {\noindent\emph{Proof.} }
               {\par}
\newcounter{example}

\def\squareforqed{\hbox{\rlap{$\sqcap$}$\sqcup$}}
\def\qed{\ifmmode\squareforqed\else{\unskip\nobreak\hfil
\penalty50\hskip1em\null\nobreak\hfil\squareforqed
\parfillskip=0pt\finalhyphendemerits=0\endgraf}\fi}

\makeatletter
\renewcommand\section{\@startsection {section}{1}{\z@}%
                                   {-1.6ex \@plus -1ex \@minus -.2ex}%
                                   {0.8ex \@plus.2ex}%
                                   {\rightskip1pc plus1fil\normalfont\normalsize\bfseries}}
\renewcommand\subsection{\@startsection{subsection}{2}{\z@}%
                                     {-1.4ex\@plus -1ex \@minus -.2ex}%
                                     {0.6ex \@plus .2ex}%
                                     {\rightskip1pc plus1fil\normalfont\normalsize\it }}
\makeatother

\newcommand{\user}[1]{\textsl{#1}}

\title{A Natural Deduction style proof system for propositional
   $\mu$-calculus and its formalization in inductive type theories}

\author{Marino Miculan}
\address{Dipartimento di Matematica e Informatica,
        Universit\`a degli Studi di Udine\\
        Via delle Scienze, 206, 33100, Udine, Italy.
        \texttt{mailto:miculan@dimi.uniud.it}}
\begin{document}
\maketitle

\abstracts{In this paper, we present a formalization of Kozen's
propositional modal $\mu$-calculus, in the Calculus of Inductive
Constructions.  We address several problematic issues, such as the use
of \emph{higher-order abstract syntax} in inductive sets in presence
of recursive constructors, the encoding of modal (``proof'') rules and
of context sensitive grammars. The encoding can be used in the \Coq\
system, providing an experimental computer-aided proof environment for
the interactive development of error-free proofs in the
$\mu$-calculus. The techniques we adopted can be readily ported to
other languages and proof systems featuring similar problematic
issues.}

\section*{Introduction}
In this paper, we present a formalization of Kozen's propositional
modal $\mu$-calculus \cite{kozen:mucalc}, often referred to as \uK, in
the \Coq\ proof assistant \cite{coq:manual}.

The $\mu$-calculus is a temporal logic which subsumes many modal and
temporal logics, such as \PDL, $CTL$, $CTL^*$, $ECTL$. Despite its
expressive power, \uK\ enjoys nice properties such as decidability and
the finite model property. The long-standing open problem of
axiomatizability of \uK\ has been solved by Walukiewicz
\cite{walu:compl,walu:notes}, who has proved the completeness of
Kozen's original system given in \cite{kozen:mucalc}.  Therefore, the
$\mu$-calculus is an ideal candidate as a logic for the verification
of processes. Nevertheless, like any formal systems, its applicability
to non trivial cases is limited by long, difficult, error-prone proofs.

This drawback can be (partially) overcome by supplying the user with a
\emph{computer-aided proof environment}, that is, a system in which he
can represent (\emph{encode, formalize}) the formal system, more or
less abstractly: its syntax, axioms, rules and inference
mechanisms. After having supplied the proof environment with a
representation of the formal system, the user should be able to
correctly manipulate (the representations of) the proofs.

However, the implementation of a proof environment for a specific
formal system is a complex, time-consuming, and daunting task.  The
environment should provide tools for checking previously hand-made
proofs; developing interactively, step-by-step, error-free proofs from
scratch; reusing previously proved properties; even, deriving
properties automatically, when feasible, freeing the user from most
unpleasant and error-prone steps. 

An alternative, and promising solution is to develop a general theory
of logical systems, that is, a \emph{Logical Framework} (LF).  A Logical
Framework is a metalogical formalism for the specification of both the
\emph{syntactic} and the \emph{deductive} notions of a wide range of
formal systems.  Logical Frameworks always provide suitable means for
representing and deal with, in the metalogical formalism, the
\emph{proofs} and \emph{derivations} of the object formal system.
Much of the implementation effort can be expended once and for all;
hence, the implementation of a Logical Framework yields a
\emph{logic-independent proof development environment}.  Such an
environment must be able to check validity of deductions in any formal
system, after it has been provided by the specification of the system
in the formalism of the Logical Framework.

In recent years, several different frameworks have been proposed,
implemented and applied to many formal systems. \emph{Type theories}
have emerged as leading candidates for Logical Frameworks.  Simple
typed $\lambda$-calculus and minimal intuitionistic propositional
logic are connected by the well-known \emph{proposition-as-types}
paradigm \cite{church:stot,debruijn:auto}.  Stronger type theories,
such as the \emph{Edinburgh Logical Framework} (ELF)
\cite{hhp:elf,ahmp}, the \emph{Calculus of Inductive Constructions}
(CIC) \cite{coq:manual} and \emph{Martin-L\"of's type theory}
(MLTT) \cite{nps:mltt}, were especially designed, or can be fruitfully
used, as a logical framework. In these frameworks, we can represent
faithfully and uniformly all the relevant concepts of the inference
process in a logical system: syntactic categories, terms, assertions,
axiom schemata, rule schemata, tactics, etc.\ via the
\emph{judgements-as-types, proofs-as-$\lambda$-terms} paradigm
\cite{hhp:elf}.  The key concept is that of \emph{hypothetico-general}
judgement \cite{martinlof:siena}, which is rendered as a type of the
dependent typed $\lambda$-calculus of the Logical Framework.  With this
interpretation, a judgement is viewed as a type whose inhabitants
correspond to proof of this judgement.

It is worthwhile noticing that Logical Frameworks based on type theory
directly give rise to proof systems in \emph{Natural Deduction style}
\cite{gentzen:ild,prawitz:nd}.  This follows directly from the fact
that the typing systems of the underlying $\lambda$-calculi are in
Natural Deduction style, and rules and proofs are faithfully
represented by $\lambda$-terms. As it is well-known, Natural Deduction
style systems are more suited to the practical usage, since they allow
for developing proofs the way mathematicians normally reason.
%\footnote{Indeed, the
%Natural Deduction style has been introduced by Gentzen because it
%``reflects as accurately as possible the actual logical reasoning
%involved in mathematical proofs'' \cite{gentzen:ild}.}

These type theories have been implemented in logic-independent systems
such as \Coq\ \cite{coq:manual}, LEGO \cite{lpt:lego}, and ALF
\cite{mn:alf}.  These systems can be readily turned into interactive
proof development environments for a specific logic: we need only to
provide the specification of the formal system (the {\em signature}),
i.e.\ a declaration of typed constants corresponding to the syntactic
categories, term constructors, judgements, and rule schemata of the
logic.  It is possible to prove, informally but rigorously, that a
formal system is correctly, \emph{adequately} represented by its
specification in the Logical Framework.  This proof usually exhibit
bijective maps between objects of the formal system (terms, formul\ae,
proofs) and the corresponding $\lambda$-terms of the encoding.

In this paper, we investigate the applicability of this approach to
the propositional $\mu$-calculus.  Due to its expressive power, we
adopt the Calculus of Inductive Constructions, implemented in the
system \Coq.  Beside its expressive power and importance in the theory
and verification of processes, the $\mu$-calculus is interesting also
for its syntactic and proof theoretic peculiarities.  These
idiosyncrasies are mainly due to a) the negative arity of ``$\mu$''
(i.e., the bound variable $x$ ranges over the same syntactic class of
$\mu x\phi$); b) context-sensitive grammar (the condition on the
formation of $\mu x\phi$); c) rules with complex side conditions
(``proof rules'').  These anomalies escape the ``standard''
representation paradigm of Logical Frameworks; that is, there is no
``standard'' way to represent them in a Logical Framework.  Hence, we
will adopt new efficient representation techniques, which can be
ported to other systems featuring the same anomalies. Moreover, since
generated editors allow the user to reason ``under assumptions'', the
designer of a proof editor for a given logic is urged to look for a
Natural Deduction formulation which can take best advantage of the
possibility of manipulating assumptions.

Beside these practical and theoretical motivations, this work can give
insights in the expressive power of CIC and \Coq. Indeed, the encoding
techniques we will adopt take full advantage of pragmatic features
offered by \Coq, such as the automatic simplification of terms, in
order to simplify as much as possible the task of proof development.

\noindent \textbf{Structure of this paper.}  In Section
\ref{sec:mucalcss}, we recall the language and the semantics of $\uK$.
We will also introduce a semantical consequence relation, which will
be the semantical counterpart of the proof system.  The Natural
Deduction style proof system $\N{\uK}$ will be introduced in Section
\ref{sec:mucalcps}. In this section we will present also a proof
system for capturing the well formedness condition on formul\ae\ of
the form $\mu x\phi$.  In Section \ref{sec:mucalcenc} we will discuss
the formalization of $\uK$ in CIC.  We will see that $\uK$ arises some
peculiarities which are difficulty encoded in CIC; we will present
some solutions.  We will suppose the reader to be familiar with the
CIC and the \Coq\ system.

Final comments and remarks are reported in Section \ref{sec:concl}.
Longer listings of \Coq\ code are reported in appendix.

\section{Syntax, semantics and consequence relation}\label{sec:mucalcss}
The language of \uK\ is an extension of the syntax of propositional
dynamic logic.  Let \ct{Act} be a set of \emph{actions} (ranged over
by $a,b,c$), $\Phi_0$ a set of atomic propositional letters (ranged
over by $p$), and \ct{Var} a set of propositional variables (ranged
over by $x,y,z$); then, the syntax of the $\mu$-calculus on \ct{Act} is:
\begin{eqnarray*}
  \Phi:\ \ \phi &::=& p \mid \mc{ff} \mid \neg\phi \mid 
                      \phi\supset\psi \mid \BOX{a}{\phi}\mid x\mid\mu x\phi
\end{eqnarray*}
where the formation of $\mu x.\phi$ is subject to the \emph{positivity
condition:} every occurrence of $x$ in $\phi$ has to appear inside an
even number of negations (In the following we will spell out this
condition more in detail).  We call \emph{preformul\ae}\ the language
obtained by dropping the positivity condition.  The variable $x$ is
\emph{bound} in $\mu x\phi$; the usual conventions about
$\alpha$-equivalence apply. We write $\nu x\phi$ as a shorthand for
$\neg\mu x(\neg\phi[\neg x/x])$.

The interpretation of $\mu$-calculus comes from Modal Logic.  A model
for the $\mu$-calculus is a transition system, that is, a pair \(\M
=\langle S,\dsb{\cdot}_{\M}\rangle\) where $S$ is a (generic) nonempty
set of \emph{(abstract) states}, ranged over by $s,t,r$, and
$\dsb{\cdot}_{\M}$ is the interpretation of atomic propositional and
command symbols: for all $p,a$, we have $\dsb{p}_\M\subset S$ and
$\dsb{a}_\M : S\to\ps(S)$.

Formul\ae\ of $\mu$-calculus may have free propositional variables;
therefore, we need to introduce \emph{environments}, which are
functions assigning sets of states to propositional variables:
$\ct{Env}\defeq\ct{Var}\to\ps(S)$.  Given a model $\M=\langle
S,\dsb{\cdot}\rangle$ and an environment $\rho$, the semantics of a
formula is the set of states in which it holds, and it is defined by
extending $\dsb{\cdot}$ compositionally, as follows:
\[
\begin{array}{@{}r@{\ \defeq\ }l@{}}
  \dsb{p}_\M\rho  & \dsb{p}\\
  \dsb{\mc{ff}}_\M\rho & \emptyset\\
  \dsb{x}_\M\rho & \rho(x)\\
  \dsb{\neg\phi}_\M\rho  & S\setminus\dsb{\phi}_\M\rho
\end{array}
\hspace{5mm}
\begin{array}{r@{\ \defeq\ }l}
  \dsb{\phi\supset\psi}_\M\rho  &
                  (S\setminus\dsb{\phi}_\M\rho)\cup\dsb{\psi}_\M\rho\\
  \dsb{\BOX{a}{\phi}}_\M\rho  &
           \{s\in S\mid \forall r\in\dsb{a}s: r\in\dsb{\phi}_\M\rho\}\\
  \dsb{\mu x\phi}_\M\rho &
      \bigcap\{T\subseteq S\mid \dsb{\phi}_\M\rho[x\mapsto T]\subseteq T\}
\end{array}
\]

It is customary to view a formula $\phi$ with a free variable $x$ as
defining a function $\phi_x^\rho:\ps(S)\to\ps(S)$, such that for all
$U\subseteq S$: $\phi_x^\rho(U) = \dsb{\phi}_\M\rho[x\mapsto U]$.  The
intuitive interpretation of $\mu x\phi$ is then the \emph{least fixed
  point} of $\phi_x^\rho$. The condition on the formation of $\mu
x\phi$ ensures the existence of the lfp:
\begin{proposition}
  Let $\phi$ a formula and $x$ a variable occurring only positively in
$\phi$.  Then, in every environment $\rho$, $\phi_x^\rho$ has both the
least and the greatest fixed point. In particular, the lfp of
$\phi_x^\rho$ is $\dsb{\mu x\phi}\rho$.
\end{proposition}
\begin{proof} (Sketch)
  It is easy to show, by induction on the syntax of $\phi$, that
  $\phi_x^\rho$ is monotone; the result follows from Knaster-Tarski's
  theorem.  \qed
\end{proof}
Notice that this result does not hold if we drop the condition on the
formation of $\mu x\phi$: for instance, the formula $\neg x$
identifies the function $(\neg x)_x^\rho(T)=S\setminus T$, which is
not monotone and has no lfp.

%\section{Consequence Relations}\label{sec:mucalccr}
In order to have a semantical counterpart of the syntactic notion of
``deduction'', we introduce a consequence relation for the
$\mu$-calculus, which is an extension of the finitary truth CR's of
propositional dynamic logic \cite{mik:eltop}.
%Definition 6.1 in  \csname b@mik:eltop\endcsname.

\begin{definition}[Consequence Relations for \uK]\label{def:mucalccr}
  Let \M\ be a model for \uK\ and \(\dsb{\cdot}_{\M}\) be the
  interpretation in \M.  The \emph{(truth) consequence relation} for
  \uK\ wrt \M\ is a relation
  \(\tcr_{\M}\subseteq\ps(\Phi)\times\Phi\), defined as follows:
  $\Gamma\tcr_\M\phi \iff
        \forall\rho.\dsb{\Gamma}_{\M}\rho\subseteq\dsb{\phi}_{\M}\rho$.

The (absolute) {\em truth CR} for \uK\ is: %defined as follows:
$\Gamma\tcr\phi \iff \forall\M.\Gamma\tcr_{\M}\phi$.

The \emph{finitary} truth consequence relations is the restriction of
$\tcr$ to finite sets:  $\Gamma\ftcr\phi \iff
     \exists \Delta\subseteq\Gamma,\mbox{\ finite}.\Delta\tcr\phi$.
\end{definition}

In the following, for sake of simplicity, we will drop the $_f$, denoting by
$\tcr$ the \emph{finitary} CR $\ftcr$.

\begin{figure}[t]
\small
\begin{eqnarray*}
\N{C}&=&
\begin{array}{|c@{\hspace{2em}}c|}\hline
   \Gamma,\phi\vdash\phi
   &
   \rl{Raa}
      {\Gamma,\neg\phi\vdash\mc{ff}}
      {\Gamma\vdash\phi}
   \\
   \rl{$\supset$-I}
      {\Gamma,\phi \vdash \psi}
      {\Gamma \vdash \phi\supset\psi}
   &
   \rl{$\supset$-E}
      {\Gamma\vdash\phi\supset\psi \spc \Gamma\vdash\phi}
      {\Gamma\vdash\psi}
   \\
   \rl{$\neg$-I}
      {\Gamma,\phi\vdash\mc{ff}}
      {\Gamma\vdash\neg\phi}
   &
   \rl{\mc{ff}-I}
      {\Gamma\vdash\phi \spc \Gamma\vdash\neg\phi}
      {\Gamma\vdash\mc{ff}}
    \\\hline
  \end{array}
\\
\N{K}&=&\N{C}+
   \begin{array}{|c@{\hspace{2em}}c|}\hline
   \rl{$[\cdot]$-I}
      {\emptyset\vdash\phi}
      {\emptyset\vdash[a]\phi}
   &
   \rl{$\supset_{[\cdot]}$-E}
      {\Gamma\vdash [a](\phi\supset\psi) \spc \Gamma\vdash [a]\phi}
      {\Gamma\vdash [a]\psi}
   \\\hline
   \end{array}
\\
\N{\uK}&=&\N{K} + \begin{array}{|c@{\hspace{2em}}c|}\hline
     \rl{$\mu$-I}
        {\Gamma\vdash\phi[(\mu x.\phi)/x]}
        {\Gamma\vdash\mu x.\phi}
        &
     \rl{$\mu$-E}
        {\Gamma\vdash\mu x.\phi \spc  \phi[\psi/x]\vdash \psi}
        {\Gamma\vdash\psi}
    \\\hline
 \end{array}
\end{eqnarray*}\vspace{-2ex}
\caption{ND-style systems for classical logic, modal logic and
  propositional $\mu$-calculus}\label{fig:nukps}
\vspace{-4ex}
\end{figure}

\section{A Natural Deduction style system for \uK}\label{sec:mucalcps}
Usually, systems for $\mu$-calculus are given in Hilbert style
\cite{kozen:mucalc,stirling:mtl,andersen:tesi}. Here  we  present  a
Natural Deduction style system for \uK, namely \N{\uK}. This system is
composed by a system for classical propositional logic (\N{C}),
extended with two rules for the minimal modal logic (\N{K}) and the
two the new two rules (introduction and elimination) for the new
constructor $\mu$, as presented in Figure \ref{fig:nukps}.  These
rules are presented in a sequent (Gentzen-like) fashion; this allow us
to spell out clearly the side conditions on hypotheses in the $\mu$-E
and $[\cdot]$-I rules. Of course, all these rules can be written also in the
more customary Natural Deduction style, like the following:
\begin{eqnarray*}
 \rl{$\mu$-E}
    {\begin{array}{cc}
              & [\phi[\psi/x]]\\
              &  \vdots\\
      \mu x.\phi &  \psi
     \end{array}}
    {\psi}\ 
 \parbox{30mm}{\small $\psi$ does not depend on any other hypothesis beside $\phi[\psi/x]$}
&\hspace{1cm}&
 \rl{$[\cdot]$-I}{\phi}{[a]\phi}\ 
 \parbox{26mm}{\small $\phi$ does not depend on any hypothesis}
\end{eqnarray*}
Notice that the side conditions of these two rules are very the same: in fact,
$\mu$-E can be stated as
$\frac{\Gamma\vdash\mu x.\phi\spc
          \emptyset\vdash \phi[\psi/x]\supset \psi}
         {\Gamma\vdash\psi}$;
here, the left subderivation has to depend on no assumptions, like to
the necessitation rule \rlref{$[\cdot]$-I} of modal logic.

The rules for $\mu$ have a direct semantic interpretation: the
introduction rule states that (the meaning of) $\mu x\phi$ is a
prefixed point of $\phi_x^\rho$; the elimination rule states that (the
meaning of) $\mu x\phi$ implies, and then ``is less than'', any
prefixed point of $\phi_x^\rho$.  Therefore, these rules state that
(the meaning of) $\mu x\phi$ is the minimum prefixed point, i.e.\ the
least fixed point, of $\phi_x^\rho$.

The resulting system is then sound and complete with respect to the
(finitary) truth consequence relation:
\begin{theorem}\label{prop:nuk}
  For $\Gamma$ finite set of formul\ae, $\phi$ formula:
  $\Gamma\vdash\phi \iff \Gamma\tcr\phi$.
\end{theorem}
\begin{proof} (Sketch)
  Soundness is proved by showing that each rule is sound.
Completeness can be proved as follows. Since $\Gamma$ is finite,
$\Gamma\tcr\phi \iff \tcr \bigwedge\Gamma\supset\phi$. By completeness
of Kozen's axiomatization \cite{walu:compl,walu:notes}, there is an
Hilbert-style derivation of $\bigwedge\Gamma\supset\phi$. Therefore,
it is sufficient to prove that Kozen's axioms and rules (e.g.\ those
presented in \cite{stirling:mtl}) are derivable in \N{\uK}.
\qed\vspace{-1ex}
\end{proof}

%\subsection{A proof system for the positivity condition}\label{sec:+-ps}
\begin{figure}\small
\[\begin{array}{|@{\ }l@{\ }|@{\ }l@{\ }|}\hline
\rl{PosinP}
   {p\in\Phi_0}
   {\mc{posin}(x,p)}
&
\rl{NeginP}
   {p\in\Phi_0}
   {\mc{negin}(x,p)}
\\
\rl{PosinY}
   {y\in\ct{Var}}
   {\mc{posin}(x,y)}
&
\rl{NeginY}
   {y\neq x}
   {\mc{negin}(x,y)}
\\
\rl{PosinImp}
   {\mc{negin}(x,\phi)\spc\mc{posin}(x,\psi)}
   {\mc{posin}(x,\phi\supset\psi)}
&
\rl{NeginImp}
   {\mc{posin}(x,\phi)\spc\mc{negin}(x,\psi)}
   {\mc{negin}(x,\phi\supset\psi)}
\\
\rl{PosinNeg}
   {\mc{negin}(x,\phi)}
   {\mc{posin}(x,\neg\phi)}
&
\rl{NeginNeg}
   {\mc{posin}(x,\phi)}
   {\mc{negin}(x,\neg\phi)}
\\
\rl{PosinBox}
   {\mc{posin}(x,\phi)}
   {\mc{posin}(x,\BOX{a}{\phi})}
&
\rl{NeginBox}
   {\mc{negin}(x,\phi)}
   {\mc{negin}(x,\BOX{a}{\phi})}
\\
\rl{PosinMu}
   {\mbox{for}\ z\neq x: \mc{posin}(x,\phi[z/y])}
   {\mc{posin}(x,\mu y\phi)}
&
\rl{NeginMu}
   {\mbox{for}\ z\neq x: \mc{negin}(x,\phi[z/y])}
   {\mc{negin}(x,\mu y\phi)}\\\hline
\end{array}\vspace{-2ex}\]
\caption{The positivity proof system.}
\label{fig:+-ps}\vspace{-4ex}
\end{figure}

Since we aim to encode the $\mu$-calculus in some logical framework,
we need to enforce the context-sensitive condition on the formation of
formul\ae\ of the form $\mu x\phi$. That is, we ought to specify in
detail the condition of ``occurring positive in a formula'' for a
variable. This notion can be represented by two new judgements on
formul\ae\ and variables, \mc{posin} and \mc{negin}, which are derived
by means of the rules in Figure \ref{fig:+-ps}.  Roughly,
$\mc{posin}(x,\phi)$ holds iff all occurrences of $x$ in $\phi$ are
positively; dually, $\mc{negin}(x,\phi)$ holds iff all occurrences of
$x$ in $\phi$ are negative. Notice that if $x$ does not occur in
$\phi$, then it occurs both positively and negatively.

Let us formalize better the meaning of these auxiliary judgements. The
notions they capture are the following:
\begin{definition}[Monotonicity and Antimonotonicity]\label{def:monoantimono}
Let $\phi$ be a formula and $x$ a variable. We say that $\phi$ is
\emph{monotone} on $x$ (written $\mc{Mon}_x(\phi)$) iff
$\forall\M,\forall\rho,\forall U,V\subseteq S$: $U\subseteq V
\Longrightarrow \phi_x^\rho(U)\subseteq\phi_x^\rho(V)$.
We say that $\phi$ is \emph{antimonotone} on $x$ (written
$\mc{AntiMon}_x(\phi)$) iff $\forall\M,\forall\rho,\forall
U,V\subseteq S$: $U\subseteq V \Longrightarrow
\phi_x^\rho(U)\supseteq\phi_x^\rho(V)$.
\end{definition}
These notions refer directly to the semantic structures in which
formul\ae\ take meaning. The following result proves that the
syntactic condition of positivity (respectively, negativity) captures
correctly the semantic condition of monotonicity (respectively,
antimonotonicity).

\begin{proposition}\label{prop:+-sound}
For all $\phi\in\Phi$, $x\in Var$:
$\vdash\mc{posin}(x,\phi) \Ra \mc{Mon}_x(\phi)$ and
$\vdash\mc{negin}(x,\phi) \Ra \mc{AntiMon}_x(\phi)$.
\end{proposition}
\begin{proof}
By simultaneous induction on the syntax of $\phi$.
%(which is equivalent to an induction on the proofs of
%$\vdash\mc{posin}(x,\phi)$, $\vdash\mc{negin}(x,\phi)$)
\qed
\end{proof}

%Proof of this proposition is given in Appendix \ref{sec:+-soundproof}.

Notice that the converse of Proposition \ref{prop:+-sound} does not
hold. Consider e.g.\ $\phi\defeq (x\supset x)$: clearly,
$\dsb{\phi}_\M^\rho=S$ always, and hence $(x\supset x)_x^\rho$ is both
monotone and antimonotone. However, $x$ does not occur only positively
nor only negatively in $\phi$. Correspondingly, we cannot derive
$\vdash\mc{posin}(x,(x\supset x))$ nor $\vdash\mc{negin}(x,(x\supset
x))$.
%: by inspection of the proof system, the only rule for deriving
%$\mc{posin}(x,(x\supset x)$ is \rlref{PosinImp}; hence, we should
%derive $\mc{negin}(x,x)$, which is not possible (rule \rlref{NeginY}).
%A similar argument applies for $\mc{negin}(x,(x\supset x))$.
This result can be generalized in the following limitation property:
\begin{proposition}
  If $x\in\fv(\phi)$ occurs both positively and negatively in $\phi$
  then neither $\mc{posin}(x,\psi)$ nor $\mc{negin}(x,\psi)$ are derivable.
\end{proposition}
\begin{proof} (Sketch)
By induction on the syntax of $\phi$.
\qed
\end{proof}
However, we can restrict ourselves to only positive formul\ae\
w.l.o.g.: by Lyndon Theorem \cite{dh:lyndon}, every monotone
formula is equivalent to a positive one.

\section{The encoding of $\mu$-calculus}\label{sec:mucalcenc}
In this section we present the encoding of the $\mu$-calculus in the
Calculus of Inductive Constructions.  We will present both the
formalization of the language and of the proof system $\N{\uK}$ given
in Section \ref{sec:mucalcps}.

\subsection{Encoding the language}
The encoding of the language of $\mu$-calculus is quite elaborate.
The customary approach, is to define an inductive type,
\texttt{o:Set}, whose constructors correspond to those of the language
of $\uK$.  In order to take full advantage of $\alpha$-conversion and
substitution machinery provided by the metalanguage, we adopt the
\emph{higher order abstract syntax} \cite{hhp:elf,dfh:hosc}.  In this
approach, binding constructors (like $\mu$) are rendered by
higher-order term constructors; that is, they take a \emph{function}.
The na{\"\i}ve representation of $\mu$, therefore, would be
\texttt{mu:(o->o)->o}; however, this solution does not work inside an
inductive definition of CIC, because it leads to a non-well-founded
definition \cite{coq:manual,dfh:hosc,mik:eltop}.

The second problem is the presence of a context-sensitive condition on
the applicability of $\mu$: in order to construct a formula of the
form $\mu x\phi$, we have to make sure that $x$ occurs positively in
$\phi$.  Inductive types do not support this kind of
restriction, since they define only context-free languages
\cite{mik:eltop}.

In order to overcome the first problems, we adopt the
  \emph{bookkeeping via Leibniz equality} technique \cite{mik:eltop}.
  We introduce a separate type, \texttt{var}, for the identifiers. The
  r\^ole played by variables is that of ``placeholders'' for
  formul\ae: they can be replaced by formul\ae\ in the application of
  $\mu$-I and $\mu$-E rules. However, we do not introduce any
  substitution predicate: instead, as we will see in Section
  \ref{sec:mucalcpsenc}, we will inherit the substitution machinery
  directly from the metalanguage, i.e., the typed $\lambda$-calculus
  of CIC.

There are no constructors for type \texttt{var}: we only assume that
there are infinitely many variables.%\vspace{-1.2ex}
\begin{verbatim}
Parameter var : Set.
Axiom var_nat : (Ex [srj:var->nat](n:nat)(Ex [x:var](srj x)=n)).
\end{verbatim}%\vspace{-1.2ex}
Then, we define the set of preformul\ae\ of $\mu$-calculus, also those
not well formed:%\vspace{-1.2ex}
\begin{verbatim}
Parameter Act : Set.
Inductive o   : Set :=  p : o | ff : o | Not : o -> o
                    | Imp : o -> o -> o
                    | Box : Act -> o -> o
                    | Var : var -> o
                    | mu  : (var->o) -> o.
\end{verbatim}%\vspace{-1.2ex}
Notice that, the argument of \texttt{mu} is a function of type
\texttt{var->o}.  In general, this may arise \emph{exotic terms},
i.e. terms which do not correspond to any preformula of the
$\mu$-calculus \cite{dfh:hosc,mik:eltop}.  In our case, this is
avoided since \texttt{var} is not declared as an inductive set (see
Section 11.2 of \cite{mik:eltop} for further details).

Now, we have to rule out all the non-well-formed formul\ae.  At the
moment, the only way for enforcing in CIC context-sensitive conditions
over languages is to define a subtype by means of $\Sigma$-types.  As
a first step, we formalize the system for positivity/negativity
presented in Figure \ref{fig:+-ps}, introducing two judgements
\texttt{posin}, \texttt{negin} of type \texttt{var->o->Prop}.  A
careful analysis of the proof system (Figure \ref{fig:+-ps}) points
out that the derivation of these judgements is completely syntax
driven. It is therefore natural to define these judgements as
\emph{recursively defined functions}, instead of inductively defined
propositions.  This is indeed possible, but the rules for the binding
operators introduce an implicit quantification over the set of
variables \emph{different from the one we are looking for}. This
quantification is rendered by assuming a locally new variable ($y$)
and that it is different from the variable $x$ (see last cases):
\begin{verbatim}
Fixpoint posin [x:var;A:o] : Prop :=
   <Prop>Case A of True
                   True
                   [B:o](negin x B)
                   [A1,A2:o](negin x A1)/\(posin x A2)
                   [a:Act][A1:o](posin x A1)
                   [y:var]True
                   [F:var->o](y:var)~(x=y) -> (posin x (F y))
         end
with negin [x:var;A:o] : Prop :=
   <Prop>Case A of True
                   True
                   [B:o](posin x B)
                   [A1,A2:o](posin x A1)/\(negin x A2)
                   [a:Act][A1:o](negin x A1)
                   [y:var]~(x=y)
                   [F:var->o](y:var)~(x=y) -> (negin x (F y))
         end.
\end{verbatim}
Therefore, in general a goal $(posin\ x\ A)$ can be
\texttt{Simpl}ified (i.e., by applying the \texttt{Simpl}
tactic, in \Coq) to a conjunction of only three forms of
propositions: \texttt{True}, negations of equalities or implications
from negations of equalities to another conjunction of the same
form. These three forms are dealt with simply in the \Coq\
environment, hence proving this kind of goals is a simple and
straightforward task.

Then, we can define when a preformula is well formed; namely, when
every application of $\mu$ satisfies the positivity condition:
\begin{verbatim}
Fixpoint iswf [A:o] : Prop :=
   <Prop>Case A of True
                   True
                   [A1:o](iswf A1)
                   [A1:o][A2:o](iswf A1)/\(iswf A2)
                   [a:Act][A1:o](iswf A1)
                   [x:var]True
                   [F:var->o](x:var)
                      ((notin x (mu F))-> (posin x (F x)))
                      /\(iswf (F x))
         end.
\end{verbatim}
Hence, each formula of the $\mu$-calculus is represented by a pair
preformula-proof of its well-formedness:
\begin{verbatim}
Record wfo : Set := mkwfo {  prp : o;  cnd : (iswf prp) }.
\end{verbatim}
%\vspace{-0.6ex}

In order to estabilish that our encoding is faithful, we introduce the
following notation: for $X=\{x_1,\ldots,x_n\} \subset Var$, let
$\Phi_X\defeq \{\phi\mid \fv(\phi)\subseteq X\}$, $\Gamma_X \defeq
\mathtt{x_1:var,\ldots,x_n:var}$; moreover, let $\texttt{o}_X \defeq
\{\texttt{t} \mid \Gamma_X\vdash\mathtt{t:o}, \mbox{ \texttt{t}
canonical}\}$ and $\texttt{wfo}_X \defeq \{\texttt{t}\in \texttt{o}_X
\mid \exists \mathtt{d}. \Gamma_X\vdash \mathtt{d}: \mathtt{(iswf\ t)}\}$.
We can then define the \emph{encoding} functions $\epsilon_X:\Phi_X
\to\mathtt{o}_X$, as follows:
\[\begin{array}{|rcl@{\ \ \ }rcl|}\hline
  \epsilon_X(\mc{ff}) &=& \tt ff    &
  \epsilon_X(\phi\supset\psi)  &=&
    \texttt{(Imp } \epsilon_X(\phi)\ \epsilon_X(\psi) \texttt{)}
  \\
  \epsilon_X(x) &=& \tt x       &
  \epsilon_X(\mu x\phi)  & =&
    \texttt{(mu [x:var]}\epsilon_{X,x}(\phi)\texttt{)} \\
  \epsilon_X(\neg\phi)  &=&
    \texttt{(Not } \epsilon_X(\phi)\texttt{)}     &
  \epsilon_X([a]\phi)  &=&
    \texttt{(Box a } \epsilon_X(\phi)\texttt{)}\\\hline
\end{array}\]
The faithfulness of our encoding is therefore stated in the following theorem:
\begin{theorem}
  For $X\subset Var$ finite, the map $\epsilon_X$ is a compositional
  bijection between $\Phi_X$ and $\mathtt{wfo}_X$.
\end{theorem}
\begin{proof} (Sketch)
  Long but not difficult inductions. First, we prove that \texttt{posin},
\texttt{negin} adequately represent the positivity/negativity proof
system. Due to its structure, it is easy to prove that the type
\texttt{(posin x A)} is inhabited by at most one canonical form (that is
to say, there is at most one way for proving that a preformula is
well-formed). Therefore, a preformula $\phi$ is a formula iff each
application of $\mu$ is valid, iff for each application of $\mu$ there
exists a (unique) witness of \texttt{posin}, iff there exists an
inhabitant of $\mathtt{(iswf}\ \epsilon_X(\phi) \mathtt{)}$.
\qed
\end{proof}

\subsection{Encoding the proof system \N{\uK}}\label{sec:mucalcpsenc}
In the encoding paradigm of Logical Frameworks, a proof system is
usually represented by introducing a \emph{proving judgement} over the
set of formul\ae, like \texttt{T:o -> Prop}.  A type \texttt{(T phi)}
should be intended, therefore, as ``$\phi$ is true''; any term which
inhabits \texttt{(T phi)} is a witness (a proof) that $\phi$ is true.
Each rule is then represented by a type constructor of \texttt{T}.  A
complete discussion of this paradigm, with an example of encoding of
$\N{C}$, can be found in \cite{hhp:elf,ahmp}.

However, in representing the proof system \N{\uK}, two difficult
issues arise: the encoding of proof rules, like $[\cdot]$-I and $\mu$-E,
and the substitution of formul\ae\ for variables in rule $\mu$-E.
These issues escape the standard encoding paradigm, so we have to
accommodate some special technique.

Actually, in the underlying theory of CIC there is no direct way for
enforcing on a premise the condition that it is a theorem (i.e.\ that
it depends on no assumptions) or, more generally, that a formula
depends only on a given set of assumptions.  The solution we adopt
exploits again the possibility provided by Logical Frameworks of
considering locally quantified premises, i.e.~general judgements in
the terminology of Martin-L{\"o}f; see \cite{ahmp:modals} for a
detailed description.

The basic proving judgement is \texttt{T:U->o->Prop}, where
\texttt{U} a set with {\em no} constructors.  Elements of \cm{U} will
be called {\em worlds} for suggestive reasons.  Each ``pure'' rule
(i.e., with no side condition), is parameterized over a generic world,
like the following (see Appendix \ref{sec:psenc} for the complete listing):
\begin{verbatim}
Axiom Imp_E : (w:U)(A,B:o)(T w (Imp A B)) -> (T w A) -> (T w B).
\end{verbatim}
Therefore, in a given world all the classical rules apply as usual.
It should be noticed, however, that we require a locally introduced
formula to be well formed. This is the case of $\supset$-I:
\begin{verbatim}
Axiom Imp_I : (w:U)(A,B:o)(iswf A) ->
                        ((T w A) -> (T w B)) -> (T w (Imp A B)).
\end{verbatim}
Indeed, it can be shown that if we allow for non-well formed
formul\ae\ in these ``negative positions'', we get easily an
inconsistent derivation.

Proof rules, on the other hand, are distinguished by \emph{local}
quantifications of the world parameter, in order to make explicit the
dependency between a conclusion and its premises.  The
$[\cdot]$-I rule is represented as follows:
\begin{verbatim}
Axiom Box_I: (w:U)(A:o)(a:Act)((w':U)(T w' A))->(T w (Box a A)).
\end{verbatim}
The idea behind the use of the extra parameter is that in making an
assumption, we are forced to assume the existence of a world, say $w$,
and to instantiate the judgement \texttt{T} also on $w$.  This
judgement then appears as an hypothesis on $w$. Hence, deriving as
premise a judgement, which is universally quantified with respect to
\cm{W}, amounts to establishing the judgement for a generic world on
which no assumptions are made, i.e.\ on no assumptions.

This idea can be suitably generalized to take care of a fixed number
of assumptions, like in rule $\mu$-E; here, the dependency between
conclusion and assumption is made evident:
\begin{verbatim}
Axiom mu_E   : (A:o)(w:U)(F:var->o)(iswf A) ->
               ((z:var)(notin z (mu F)) -> (Var z)=A ->
                       (w':U)(T w' (F z)) -> (T w' A))
               -> (T w (mu F)) -> (T w A).
\end{verbatim}
This is the most complex rule of the whole system: besides the world
parameter technique, it leads us to the second problematic issue of
\N{\uK}, namely the substitution of formul\ae\ for variables, by means
of Leibniz equality $=$.  A similar, but simpler situation, arises in
the encoding of $\mu$-I:
\begin{verbatim}
Axiom mu_I   : (A:o)(w:U)(F:var->o)
               ((z:var)(notin z (mu F)) -> (Var z)=(mu F)
                       -> (T w (F z)))
               -> (T w (mu F)).
\end{verbatim}
The idea is to do not perform substitution immediately; instead it is
delayed, until it is actually needed.  The binding between the
substituted variable \texttt{z} and the formula \texttt{(mu F)} is
kept in the derivation environment by the hypothesis \texttt{(Var
  z)=(mu F)}.  Moreover, this hypothesis can be used by the
\texttt{Rewrite} tactic of \Coq, for replacing automatically the
variable.  Therefore, we do not need to implement any explicit
mechanism for substitution: it is directly inherited from the
$\beta$-reduction of the underlying $\lambda$-calculus.  For an
example of application, see Section \ref{sec:example}.

We need also to locally assume the fact that $z$ does not appear in
the formula, i.e.\ it is \emph{fresh}.  This is achieved by the
hypothesis \texttt{(notin z (mu F))}.  The judgement \texttt{notin}
(and the dual \texttt{isin}, see Section \ref{sec:syntenc}) are
auxiliary judgements for occur-checking.  Roughly, \texttt{(notin x
  A)} holds iff \texttt{x} does not occur free in \texttt{A}; dually
  for \texttt{isin}.  They may be needed in the rest of derivation for
  inferring well-formness of discharged formul\ae\ in rules
  \rlref{Raa}, \rlref{$\supset$-I}, \rlref{$\neg$-I}.

The formalization of \N{\uK} we have presented is \emph{adequate},
that is, we can derive a property in the system \N{\uK} iff we can
inhabit the corresponding type in our encoding.  This is stated
precisely by the following result:
\begin{theorem}
  For $X\subset Var$ finite, for $\phi_1,\ldots,\phi_n,\phi\in\Phi_X$:
  $\phi_1,\ldots,\phi_n\vdash_{\N{\uK}}\phi$ iff
  $\exists\mathtt{t}.\Gamma_X,\mathtt{w:U}, \mathtt{a_1:(T\ w\
  }\epsilon_X(\phi_1)),\ldots, \mathtt{a_n:(T\ w\
  }\epsilon_X(\phi_n))\vdash \mathtt{t:(T\ w\ }\epsilon_X(\phi))$
\end{theorem}
\begin{proof} ($\Ra$) by induction on derivation; ($\Leftarrow$)
by induction on \texttt{t}.\qed
\end{proof}

\section{Conclusions}\label{sec:concl}
In this paper we have presented an original encoding of the
$\mu$-calculus in type-theory based logical frameworks.  We have
addressed several problematic issues. First, the extensive and wise
use of the higher order abstract syntax frees us from a tedious
encoding of the mechanisms involved in the handling of bound names
%(i.e. substitution and $\alpha$-conversion)
because they are automatically inherited from the metalevel. Secondly,
we have faithfully represented the (context-sensitive) language of
$\mu$-calculus by formalizing the notion of ``well formed formula''.
Thirdly, the modal nature of the rules of $\mu$-calculus has been
rendered, although Logical Frameworks do not support directly modal
rules.

The techniques we have adopted can be readily ported to other
formalisms featuring similar problematic issues, such
as the $\lambda$-calculus, higher-order process calculi, languages
defined by context-sensitive grammars, modal logics\ldots

Moreover, our experience confirmed also in dealing with the
$\mu$-calculus, is that Logical Frameworks allow to encode faithfully
the formal systems under consideration, without imposing on the user
of the proof editor the burden of cumbersome encodings.  However,
nowadays proof editors and Logical Frameworks are still under
development; hence, they will benefit from extensive case studies and
applications, like the one presented here, which can enlighten weak
points and suggest further improvements.

Finally, the encoding presented in this paper could be used as the
kernel for a user friendly computer-aided proof environment, in which
the user can carry out interactively formal verifications based on the
$\mu$-calculus.

\appendix

\small
\section{\Coq\ code}\label{sec:coq-code}
\subsection{Code of the encoding of the syntax}\label{sec:syntenc}
\begin{verbatim}
(* Sets for actions, variables *)
Parameter Act, var : Set.
(* var is at least enumerable *)
Axiom var_nat : (Ex [srj:var->nat](n:nat)(Ex [x:var](srj x)=n)).
Lemma neverempty : (x:var)(Ex [y:var]~(x=y)).
(* proof omitted *)

(* the set of preformulae, also not well formed *)
Inductive o   : Set := p  : o
                    | ff  : o
                    | Not : o -> o
                    | Imp : o -> o -> o
                    | Box : Act -> o -> o
                    | Var : var -> o
                    | mu  : (var->o) -> o.

Fixpoint isin [x:var;A:o] : Prop :=
   <Prop>Case A of False
                   False
                   [B:o](isin x B)
                   [A1,A2:o](isin x A1)\/(isin x A2)
                   [a:Act][B:o](isin x B)
                   [y:var]x=y
                   [F:var->o](y:var)(isin x (F y))
         end.
Fixpoint notin [x:var;A:o] : Prop :=
   <Prop>Case A of True
                   True
                   [B:o](notin x B)
                   [A1,A2:o](notin x A1)/\(notin x A2)
                   [a:Act][B:o](notin x B)
                   [y:var]~(x=y)
                   [F:var->o](y:var)~(x=y) -> (notin x (F y))
         end.

Fixpoint posin [x:var;A:o] : Prop :=
   <Prop>Case A of True
                   True
                   [B:o](negin x B)
                   [A1,A2:o](negin x A1)/\(posin x A2)
                   [a:Act][A1:o](posin x A1)
                   [y:var]True
                   [F:var->o](y:var)~(x=y) -> (posin x (F y))
         end
with negin [x:var;A:o] : Prop :=
   <Prop>Case A of True
                   True
                   [B:o](posin x B)
                   [A1,A2:o](posin x A1)/\(negin x A2)
                   [a:Act][A1:o](negin x A1)
                   [y:var]~(x=y)
                   [F:var->o](y:var)~(x=y) -> (negin x (F y))
         end.

Fixpoint iswf [A:o] : Prop :=
   <Prop>Case A of True
                   True
                   [A1:o](iswf A1)
                   [A1:o][A2:o](iswf A1)/\(iswf A2)
                   [a:Act][A1:o](iswf A1)
                   [x:var]True
                   [F:var->o](x:var)
                       ((notin x (mu F)) -> (posin x (F x)))
                       /\(iswf (F x))
         end.

(* the set of well formed formulae *)
Record wfo : Set := mkwfo { prp : o;  cnd : (iswf prp) }.

(* separation: if x does not apper in A and y do, then x and y are
 * not the same identifiers - proof omitted *)
Lemma separation : (x,y:var)(A:o)(notin x A) -> (isin y A) -> ~(x=y).

(* an identifier which does not occur,
 * occurs both positively and negatively - proof omitted *)
Lemma notin_posin_negin : 
                 (x:var)(A:o)(notin x A) -> (posin x A)/\(negin x A).
\end{verbatim}
\subsection{Code of the encoding of the proof system}\label{sec:psenc}
\begin{verbatim}
(* the universe, for the world technique *)
Parameter U : Set.

(* the proving judgement *)
Parameter T : U -> o -> Prop.

Section Proof_Rules.
Variable A,B: o.
Variable w  : U.

(* proof rules operate also on non-well formed formulae, but for
   ensuring the soundness of the system, we need to require
   well-formness of every discharged formula *)

Axiom ff_I   : (T w A) -> (T w (Not A)) -> (T w ff).
Axiom Not_I  : (iswf A) -> ((T w A) -> (T w ff)) -> (T w (Not A)).
Axiom RAA    : (iswf A) -> ((T w (Not A)) -> (T w ff)) -> (T w A).

Axiom Imp_I  : (iswf A) -> ((T w A) -> (T w B)) -> (T w (Imp A B)).
Axiom Imp_E  : (T w (Imp A B)) -> (T w A) -> (T w B).

Axiom Box_I  : (a:Act) ((w':U)(T w' A)) -> (T w (Box a A)).
Axiom K : (a:Act) (T w (Box a (Imp A B)))
          -> (T w (Box a A)) -> (T w (Box a B)).

Axiom mu_I   : (F:var->o)
             ((z:var)(notin z (mu F)) -> (Var z)=(mu F) -> (T w (F z)))
               -> (T w (mu F)).
Axiom mu_E   : (F:var->o)(iswf A) ->
               ((z:var)(notin z (mu F)) -> (Var z)=A ->
                       (w':U)(T w' (F z)) -> (T w' A))
               -> (T w (mu F)) -> (T w A).
End Proof_Rules.

Lemma ff_E : (A:o)(iswf A) -> (w:U)(T w ff) -> (T w A).
Intros; Apply RAA; Intros; Assumption.
Qed.
\end{verbatim}
\subsection{An example session in \Coq}\label{sec:example}
We will show a complete \Coq\ session, in which we prove that
$A\supset\mu x(A\supset x)\vdash \mu x(A\supset x)$.  Commands entered
by the user are written in \texttt{\user{this font}}.
\begin{alltt}
miculan@maxi:~> \user{coqtop}
Welcome to Coq V6.2 (May 1998)
Coq < \user{Require mu.}
[Reinterning mu ...done]
\end{alltt}
Let $w$ be a world and $A$ a formula:
\begin{alltt}
Coq < \user{Variable w:U. Variable A:o.}
w is assumed
A is assumed
\end{alltt}
We claim the lemma we intend to prove; this leads us in ``proof mode'':
\begin{alltt}
Coq < \user{Lemma simple : (T w (Imp A (mu [x:var](Imp A (Var x))))) ->}
                     \user{(T w (mu [x:var](Imp A (Var x)))).}
1 subgoal
  ============================
   (T w (Imp A (mu [x:var](Imp A (Var x)))))
    ->(T w (mu [x:var](Imp A (Var x))))

simple < \user{Intros.}
1 subgoal
  H : (T w (Imp A (mu [x:var](Imp A (Var x)))))
  ============================
    (T w (mu [x:var](Imp A (Var x))))

simple < \user{Apply mu\_I; Intros.}
1 subgoal
  H : (T w (Imp A (mu [x:var](Imp A (Var x)))))
  z : var
  H0 : (notin z (mu [x:var](Imp A (Var x))))
  H1 : (Var z)=(mu [x:var](Imp A (Var x)))
  ============================
   (T w (Imp A (Var z)))
\end{alltt}
Now, we need to replace \texttt{z} by the corresponding formula, in
order to conclude:
\begin{alltt}
simple < \user{Rewrite H1.}
1 subgoal
  H : (T w (Imp A (mu [x:var](Imp A (Var x)))))
  z : var
  H0 : (notin z (mu [x:var](Imp A (Var x))))
  H1 : (Var z)=(mu [x:var](Imp A (Var x)))
  ============================
   (T w (Imp A (mu [x:var](Imp A (Var x)))))

simple < \user{Apply H.}
Subtree proved!

simple < \user{Qed.}
(Intros; Apply mu\_I; Intros).
Rewrite H1.
Apply H.
simple is defined
\end{alltt}

\bibliographystyle{abbrv}
\bibliography{allbib}

\begin{thebibliography}{10}

\bibitem{hlics}
\newblock {\em Handbook of Logic in Computer Science}.
\newblock Oxford University Press, 1992.

\bibitem{ahmp}
A.~Avron, F.~Honsell, I.~A. Mason, and R.~Pollack.
\newblock Using {T}yped {L}ambda {C}alculus to implement formal systems on a
  machine.
\newblock {\em Journal of Automated Reasoning}, 9:309--354, 1992.

\bibitem{ahmp:modals}
A.~Avron, F.~Honsell, M.~Miculan, and C.~Paravano.
\newblock Encoding modal logics in {L}ogical {F}rameworks.
\newblock {\em Studia Logica}, 60(1):161--208, Jan. 1998.

\bibitem{coq:manual}
\newblock {\em The Coq Proof Assistant Reference Manual - Version 6.2}.
\newblock INRIA, Rocquencourt, May 1998.
\newblock Available at \texttt{ftp://ftp.inria.fr/INRIA/coq/V6.2/doc}.

\bibitem{church:stot}
A.~Church.
\newblock A formulation of the simple theory of types.
\newblock {\em JSL}, 5:56--68, 1940.

\bibitem{dh:lyndon}
G.~D'Agostino, M.~Hollenberg.
\newblock Logical questions concerning the $\mu$-calculus:
interpolation, Lyndon, and {\L}o\'s-Tarski.
\newblock To be published in the \emph{JSL}, 1999.

\bibitem{dfh:hosc}
J.~Despeyroux, A.~Felty, and A.~Hirschowitz.
\newblock Higher-order syntax in {C}oq.
\newblock In {\em Proc.\ of TLCA'95}. LNCS 902, Springer-Verlag, 1995.

\bibitem{gentzen:ild}
G.~Gentzen.
\newblock Investigations into logical deduction.
\newblock In M.~Szabo, ed., {\em The collected papers of Gerhard Gentzen},
  pages 68--131. North Holland, 1969.

\bibitem{hhp:elf}
R.~Harper, F.~Honsell, and G.~Plotkin.
\newblock A framework for defining logics.
\newblock {\em J.~ACM}, 40(1):143--184, Jan. 1993.

\bibitem{kozen:mucalc}
D.~Kozen.
\newblock Results on the propositional $\mu$-calculus.
\newblock {\em TCS} 27, 1983.

\bibitem{lpt:lego}
Z.~Luo, R.~Pollack, and P.~Taylor.
\newblock {\em How to use {LEGO}}.
\newblock Department of Computer Science, University of Edinburgh, Oct. 1989.

\bibitem{mn:alf}
L.~Magnusson and B.~Nordstr{\"o}m.
\newblock The {ALF} proof editor and its proof engine.
\newblock In {\em Proc. of  TYPES'93}, LNCS 806, pages 213--237.
  Springer-Verlag, 1994.

\bibitem{martinlof:siena}
P.~{Martin-L\"of}.
\newblock On the meaning of the logical constants and the justifications of the
  logic laws.
\newblock TR~2, Dipartimento di Matematica, Universit\`a di Siena, 1985.

\bibitem{mik:eltop}
M.~Miculan.
\newblock {\em Encoding Logical Theories of Programs}.
\newblock PhD thesis, Dipartimento di Informatica, Universit\`a di Pisa, Italy,
  Mar. 1997.

\bibitem{mik:mucode}
M.~Miculan.
\newblock \Coq\ signature for the $\mu$-calculus.
\newblock Available at \texttt{http://www.dimi.uniud.it/\~{ }miculan/mucalculus}

\bibitem{nps:mltt}
B.~Nordstr{\"o}m, K.~Petersson, and J.~M. Smith.
\newblock Martin-{L\"o}f's type theory.
\newblock In \cite{hlics}.

\bibitem{prawitz:nd}
D.~Prawitz.
\newblock {\em Natural Deduction}.
\newblock Almqvist \& Wiksell, Stockholm, 1965.

\bibitem{stirling:mtl}
C.~Stirling.
\newblock Modal and {T}emporal {L}ogics.
\newblock In \cite{hlics}, pages 477--563.

\bibitem{walu:compl}
I.~Walukiewicz.
\newblock Completeness of {Kozen}'s axiomatisation.
\newblock In D.~Kozen, editor, {\em Proc.\ 10th LICS}, pages 14--24, June 1995. IEEE.

\end{thebibliography}
\vspace{\fill}
\noindent \textbf{Acknowledgements.}
The author is grateful to an anonymous referee for many useful
remarks.
\end{document}